\documentclass[10pt,conference]{IEEEtran}

\usepackage[dvips]{graphics}
\usepackage{epsfig}
\newtheorem{assume}{Assumption}
\IEEEoverridecommandlockouts   

\begin{document}
\title{An Improved Analytic Expression for Write Amplification in NAND Flash}
\author{\IEEEauthorblockN{Xiang Luojie, Brian M. Kurkoski}
\IEEEauthorblockA{
Dept. of Information and Communications Engineering\\
The University of Electro-Communications\\
1-5-1 Chofugaoka, Chofu\\
Tokyo, Japan\\
}
\thanks{This research was supported in part by the Ministry of Education, Science, Sports and Culture; Grant-in-Aid for Scientific Research (C) number 21560388 and Grant-in-Aid for Scientific Research (C) number 23560439. Contact email: kurkoski.ice.uec.ac.jp}%

}

\maketitle
\date{}

\begin{abstract}
\boldmath Agarwal et al. gave an closed-form expression for write amplification in NAND flash memory by finding the probability of a page being valid over the whole flash memory. This paper gives an improved analytic expression for write amplification in NAND flash memory by finding the probability of a page being invalid over the block selected for garbage collection. The improved expression uses Lambert W function. Through asymptotic analysis, write amplification is shown to depend on overprovisioning factor only, consistent with the previous work. Comparison with numerical simulations shows that the improved expression achieves a more accurate prediction of write amplification. For example, when the overprovisioning factor is 0.3, the expression proposed by this paper gives a write amplification of 2.36 whereas that of the previous work gives 2.17, when the actual value is 2.35.
\end{abstract}

\section{Introduction}
Flash memory is a storage medium with growing significance. It has many appealing features including non-volatility, small size, low-cost, mechanical reliability, low power consumption and low read latencies particularly when compared to hard disk drives \cite{jagmohan}~\cite{marcus}.

Flash memory is organised in blocks. A block has a fixed number of pages (typically 64 pages). A page has a fixed size (typically 4KiB) \cite{xiaoyu}. There are three kinds of operations on flash memory: read, write and erase. Read and write operation can be performed on a page basis \cite{haas}. 

Flash memory has limitations that challenge the design of flash memory systems. One fundamental limitation of flash memory is  nonsupport of overwriting and block erase. After data are written into a page, new data can not be written into the page by overwriting. The page must be erased before new data can be written to it. But erase operation can be performed only on a block basis \cite{kawaguchi}. Another fundamental limitation of flash memory is limited endurance. Flash memory can tolerate a limited number of program and erase cycles before it becomes unreliable. The latest multi-level cell (MLC) memories can endure $5000\sim10000$ program and erase cycles \cite{jagmohan} \cite{marcus}. Thus, it is a critical problem to limit the number of program and erase cycles to a minimum to increase the flash memory's lifetime.

  Erasing an entire block when a page is needed to be updated is inefficient. Moreover, this can wear out the flash memory very quickly. Thus, in flash memory, out-of-place write is used \cite{marcus}. When a page already written needs to be updated, the new data are not written into that page but to a new page free for writing and the old page is marked as invalid. A mapping table is maintained to record the mappings of logical address and physical address. In this paper, pages are classified into three categories and their definition is given as follows. A \emph{free page} is a page into which no data have been written. It is available to accommodate new user writes. A \emph{valid page} is a page into which data have been written but hasn't been updated. A valid page is the page that stores the user data. An \emph{invalid page} is a page into which data have been written and has been updated. An invalid page once stores user data, but it is updated and no longer stores user data due to out-of-place write. 

  Out-of-place results in invalid pages. Invalid pages consume the flash memory but do not store user data. Thus, when the number of invalid pages accumulate to some extent, the invalid pages should be reclaimed for new user writes. The mechanism in flash memory to reclaim invalid pages and translate them into free pages is called garbage collection \cite{marcus}. Garbage collection is performed in the following way: first a block is selected for garbage collection. Then, the valid pages in the selected block are copied to some other free space in the flash memory. After this, the selected block is erased and becomes free. Various algorithms for garbage collection have been proposed in previous work \cite{rosenblum} \cite{menon} \cite{chang}.

  During garbage collection, valid pages in the selected block are copied to some other free space before the selected block is erased and copied back to the flash memory afterwards. This copy operation causes additional writes. The actual number of writes on the flash memory would be more than the number of pages needed to be written. This phenomenon is called write amplification. Write amplification reduces the flash memory's lifetime and therefore should be minimized. In flash memory, a common practice is that, user can only use a portion of the raw flash memory space. The portion the user cannot use is called overprovisioning. Overprovisioning provides flash memory with an increased endurance and an improved performance. Increasing the amount of overprovisioning decreases write amplification.

Due to the significance of minimizing write amplification, previous works analyzed write amplification. Hu et al. first developed a probabilistic model to describe the write amplification \cite{xiaoyu}. They then developed an empirical model to compute the write amplification \cite{haas}. 

Agarwal et al. gave a closed-form expression for write amplification \cite{marcus}. They assumed that initially, the user space of the flash memory is full and the valid pages are randomly distributed over the physical flash memory. The number of valid pages in an arbitrary block is then binomially distributed. After a sufficiently large number of user writes, the distribution of the number of valid pages in an arbitrary block is empirically approximated by a uniform distribution. Because both the total number of pages and the total number of valid pages remains the same, the average number of valid pages in an arbitrary block remains the same. Thus, the expected value of these two distributions is equal. By solving this equation, an expression for write amplification is obtained which is a function of overprovisioning factor.

This paper follows this work, and proposes an improved analytical expression for write amplification in NAND flash memory. First, $X$, the number of invalid pages freed by each garbage collection is studied. After a sufficiently large number of user writes, $X$ evolves into a stationary state. We assume that, in the stationary state, the expected value of $X$ is a constant $x$. Another assumption is made and justified that, $X$ has a binomial distribution with probability $p_{\mathsf{ie}}$. However, $p_{\mathsf{ie}}$ and the expected value of the binomial distribution depend on $x$. Since the expected value of the binomial distribution is $x$, we form an equality which can be solved for $x$. An asymptotic value of $x$ and an asymptotic value of write amplification $\widehat{A}$ are obtained using Lambert W function when the total number of pages accessible to the user is very large. As with the previous work, $\widehat{A}$ is a function of overprovisioning factor $\rho$. 

The key of both papers is finding the probability of a page being invalid or valid. Agarwal et al. find the probability of a page being valid over the whole flash memory while this paper finds the probability of a page being invalid over the block selected for garbage collection. Comparison shows that the improved expression achieves a more accurate prediction of write amplification than the previous work. For example, when the overprovisioning factor is 0.3, the improved expression gives a write amplification of 2.36 and that of the previous work gives 2.17 whereas the actual value is 2.35.
 
The rest of the paper is organized as follows: Section \ref{model} specifies the system model assumed. The analysis of write amplification and derivation of closed-form expression are given in Section \ref{analysis}. In Section \ref{simulation}, the assumptions in Section \ref{analysis} is justified and the comparison between the result of this paper and previous work is given.

\section{System Model}\label{model}
The flash memory is organized in blocks. There are a total of $T$ physical blocks in the flash memory. Each block has a fixed number of pages, denoted by $N_{\mathsf{p}}$. The portion of the flash memory space that user can use is $U$ blocks. Overprovisioning factor $\rho$ is defined by:
\begin{equation}\label{overprovisioning}
\rho=\frac{T-U}{U}.
\end{equation}

The flash memory maintains a free block pool and an occupied block queue. Blocks that have at least one free page are in the free block pool and blocks with no free pages are in the occupied block queue. Initially, the flash memory is empty and all the blocks are empty and are in the free block pool. This paper assumes random writes which are uniformly distributed on the user address space. Each write request has a fixed size of one page. When a write request comes, data are written to a free page in a block in the free block pool. If the write request falls on a user address that has not been written before, this write request is called a ``write''. If the write request falls on a user address that has already been written, this write request is called an ``update". When it is a write, a free page is found and the data are written into that page. The mapping between logical address and physical page address is recorded in a mapping table. When it is an update, the physical page storing the old data is found through the mapping table and marked as invalid. A free page would be found to write the new data and the mapping table is updated accordingly. Gradually, the blocks in the free block queue become full and are moved into the occupied block queue. When the number of free pages reaches a predefined threshold, garbage collection is triggered to collect the invalid pages. Greedy garbage collection is assumed where garbage collection is triggered when no free page exists. The block with the maximum number of invalid pages is selected for garbage collection. The valid pages in the selected block is copied to some auxiliary space in the flash memory. Then the selected block is erased. Finally the valid pages is copied back to this block. This block now joins the free block pool and is the only block in the free block pool accommodating write requests. 

Suppose there are $N_{\mathsf{ip}}$ invalid pages freed by garbage collection. Notice that, copying the valid pages contributes $N_{\mathsf{p}}-N_{\mathsf{ip}}$ writes on the flash memory. And as a result, $N_{\mathsf{ip}}$ free pages are obtained and therefore $N_{\mathsf{ip}}$ new user writes can be accommodated. Write amplification, $A$, is defined as the average number of physical page writes per user page write \cite{marcus}. It is given by:
\begin{equation}\label{define_write_amplification}
A=\frac{N_{\mathsf{p}}-N_{\mathsf{ip}}+N_{\mathsf{ip}}}{N_{\mathsf{ip}}}=\frac{N_{\mathsf{p}}}{N_{\mathsf{ip}}}.
\end{equation}

Due to the greedy garbage collection policy, at most one block can have free pages and be in the free block pool. Thus, the flash memory model can be further simplified into a single block queue. Blocks are sorted by their time in the queue since they were last selected for garbage collection. They are labeled $0,1,2,...,T-1$ from the head of the queue. When no free pages exist in any block, the block with the maximum number of invalid pages is chosen and removed from the queue. All blocks after this block moves one step towards the head of the queue. After the erase operation, this selected block is added to the tail of the queue with $N_{\mathsf{ip}}$ free pages and $N_{\mathsf{p}}-N_{\mathsf{ip}}$ valid pages. In this model, only the last block (block $T-1$) can have free pages and garbage collection is triggered when this block is full. 

\begin{figure}[h]
\centering
\scalebox{0.4}{\includegraphics{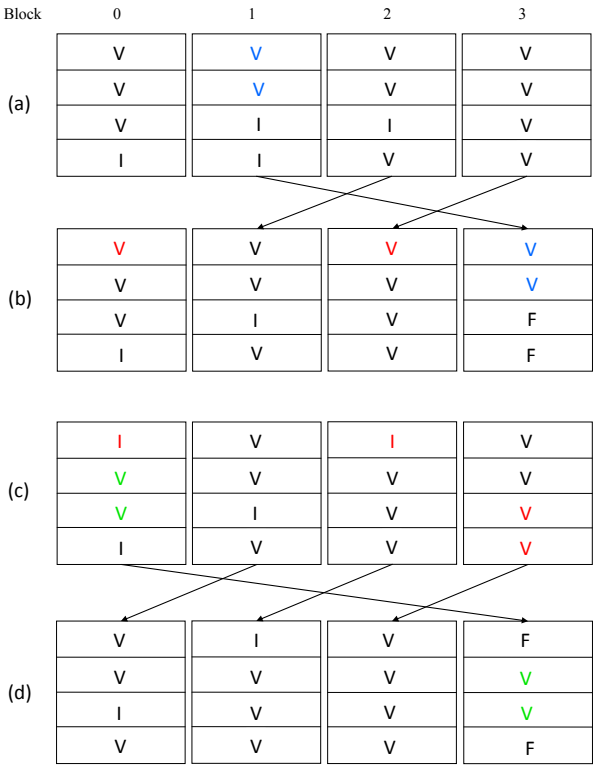}}
\caption{Demonstration of simplified model}
\label{fig:block_queue}
\end{figure}

A demonstration of this model is shown in Fig. \ref{fig:block_queue}. ``V" stands for valid, ``I" for invalid and ``F" for free. The demonstration includes four blocks labeled block 0,1,2,3 and each block has four pages. The overprovisioning factor is $\frac{1}{3}$. This means the user can only use 12 pages out of the 16 pages. In step (a), there are 12 valid pages in total indicating that the user address space is already full and only update can be performed. At this time the flash memory is full and garbage collection is triggered. Block 1 should be selected for garbage collection because it has the maximum number of invalid pages. Block 1 is removed from the queue and block 2,3 moves one step towards the head of the queue becoming block 1,2. The valid pages in the selected block shown as blue V's in step (a) are copied to some auxiliary space and then the selected block is erased and added to the tail of the queue becoming block 3 as is shown in step (b). The valid pages are copied into Block 3 shown as blue V's in step (b). The original two invalid pages now becomes free pages  accommodating new user writes. Two user writes updates the red pages shown in (b). These pages are first marked as invalid shown as red I's in (c) and new data are written into block 3 shown as red V's in (c). Now garbage collection is triggered again. Block 0 should be selected for garbage collection which is performed in the same way as in step (a). 

\section{Write Amplification Analysis}\label{analysis}
For the convenience of explanation later, first the definition of a block cycle is given. A block cycle is the process from when garbage collection is performed on a block to the time when this block is selected again for garbage collection. From the point of view of the block queue, a block cycle is the process from a block entering the block queue being block $T-1$ to the time when it is removed from the block queue for garbage collection.

Let $X$ denote the number of invalid pages freed from garbage collection. The following assumption was used in \cite{marcus} and is further supported by simulation in Section \ref{simulation}.
\begin{assume}\label{stationary_assumption}
After a sufficiently large number of user writes, $X$  evolves into a stationary state and has a constant expected value $\mathsf{E}(X)=x$.
\end{assume}

For any block selected for garbage collection, the following assumption is given and will be justified later.
\begin{assume}\label{binomial_assumption}
When a block is selected for garbage collection, all the pages in this block have the same probability $p_{\mathsf{ie}}$ to be invalid.
\end{assume}

Due to this assumption, $X$ is a binomial random variable with $N_{\mathsf{p}}$ trials and each trial has a probability  $p_{\mathsf{ie}}$ to be successful. The expected value of $X$ is $N_{\mathsf{p}}p_{\mathsf{ie}}$. Therefore, using Assumption \ref{stationary_assumption},
\begin{equation}\label{equivalent}
N_{\mathsf{p}}p_{\mathsf{ie}}=x.
\end{equation}

The following derives $p_{\mathsf{ie}}$. Consider a certain valid page $\alpha$. It corresponds to one page of data in the user space. Because the user writes are randomly distributed on the user space and are mutually independent and the total user space is $UN_{\mathsf{p}}$ pages, a user write has a probability of $\frac{1}{UN_{\mathsf{p}}}$ to update the data page $\alpha$ stores, and make it invalid. After $k$ user writes, page $\alpha$ remains valid if none of these $k$ user writes update the data that page $\alpha$ stores. Let $p_{\mathsf{valid}}$ denote the probability that a valid page $\alpha$ remains valid after $k$ user writes,
\begin{equation}\label{p_valid}
p_{\mathsf{valid}}=(1-\frac{1}{UN_{\mathsf{p}}})^k.
\end{equation}
Then, $p_{\mathsf{invalid}}$, the probability that page $\alpha$ becomes invalid after $k$ user writes:
\begin{equation}\label{p_invalid}
p_{\mathsf{invalid}}=1-p_{\mathsf{valid}}=1-(1-\frac{1}{UN_{\mathsf{p}}})^k.
\end{equation}

Let $k_{\mathsf{e}}$ denote the total number of user writes during an entire block cycle. According to Assumption \ref{binomial_assumption}, $p_{\mathsf{ie}}$ is the value of $p_{\mathsf{invalid}}$ when $k=k_{\mathsf{e}}$. Thus, to determine $p_{\mathsf{invalid}}$,  $k=k_{\mathsf{e}}$ should be determined. According to Assumption \ref{stationary_assumption}, each garbage collection collects $x$ invalid pages. This means, after each garbage collection, there are $x$ free pages in the block $T-1$ translated from invalid pages freed by garbage collection. Then, $x$ user writes fills this block and triggers the next garbage collection. Thus, between two garbage collections, one block (block $T-1$) becomes full and there are $x$ user writes. 
\begin{assume}\label{T_assumption}
A number of $T$ blocks become full during an entire block cycle.
\end{assume}

This assumption will be justified by simulation in Section \ref{simulation}. Using this assumption, 
\begin{equation}\label{k_e}
k_{\mathsf{e}}=Tx.
\end{equation}

 Apply equation (\ref{k_e}) to equation (\ref{p_invalid}), $p_{\mathsf{ie}}$, the value of $p_{\mathsf{invalid}}$ when $k=k_{\mathsf{e}}$, is:
\begin{equation}\label{p_invalid_TX}
p_{\mathsf{ie}}=1-(1-\frac{1}{UN_{\mathsf{p}}})^{Tx}.
\end{equation}

Apply equation (\ref{p_invalid_TX}) to equation (\ref{equivalent}), 
\begin{equation}\label{equation}
N_{\mathsf{p}}(1-(1-\frac{1}{UN_{\mathsf{p}}})^{Tx})=x,
\end{equation}
where $N_{\mathsf{p}}, T, U$ are constants for a specific flash memory. Let $\mathsf{W}(x)$ denote the Lambert W function \cite{lambert}. Solving this equation for $x$,
\begin{equation}\label{solve_X}
x=-\frac{\mathsf{W}(N_{\mathsf{p}}T(1-\frac{1}{UN_{\mathsf{p}}})^{TN_{\mathsf{p}}}\ln(1-\frac{1}{UN_{\mathsf{p}}}))}{T\ln(1-\frac{1}{UN_{\mathsf{p}}})}+N_{\mathsf{p}}.
\end{equation}
Thus, write amplification $A$ is,
\begin{eqnarray}\label{real_write_amplification}
A&=&\frac{N_{\mathsf{p}}}{x}\nonumber\\
&=&N_{\mathsf{p}}T\ln(1-\frac{1}{UN_{\mathsf{p}}})/(N_{\mathsf{p}}T\ln(1-\frac{1}{UN_{\mathsf{p}}})-
\nonumber\\&&\mathsf{W}(N_{\mathsf{p}}T(1-\frac{1}{UN_{\mathsf{p}}})^{TN_{\mathsf{p}}}\ln(1-\frac{1}{UN_{\mathsf{p}}}))).
\end{eqnarray}
As was mentioned, practical flash memory has a very large number of blocks so that $UN_{\mathsf{p}}\gg 1$. Consider $\widehat{x}$, the asymptotic value of $x$ as $UN_{\mathsf{p}}$ becomes large:
\begin{eqnarray}\label{practical_x}
\widehat{x}&=&\lim_{UN_{\mathsf{p}}\to\infty}{-\mathsf{W}(N_{\mathsf{p}}T(1-\frac{1}{UN_{\mathsf{p}}})^{TN_{\mathsf{p}}}\ln(1-\frac{1}{UN_{\mathsf{p}}}))}
\nonumber\\
&&/(T\ln(1-\frac{1}{UN_{\mathsf{p}}}))+N_{\mathsf{p}} \nonumber\\
&=&N_{\mathsf{p}}(1-\frac{\mathsf{W}((-1-\rho)e^{-1-\rho})}{-1-\rho}).
\end{eqnarray}
The asymptotic write amplification $\widehat{A}$ is:
\begin{eqnarray}\label{final}
\widehat{A}&=& \frac{N_{\mathsf{p}}}{\widehat{x}}\nonumber\\
&=&\frac{-1-\rho}{-1-\rho-\mathsf{W}((-1-\rho)e^{-1-\rho})}.
\end{eqnarray}
For comparison, the closed-form expression for write amplification $A_{\mathsf{pre}}$ proposed in the previous work \cite{marcus} is:
\begin{equation}\label{previous_work}
A_{\mathsf{pre}}=\frac{1}{2}\big(\frac{1+\rho}{\rho}\big).
\end{equation}
Comparison of equation (\ref{final}) and equation (\ref{previous_work}) shows that, the asymptotic write amplification expression obtained by this analysis is a function of overprovisioning factor only, which is consistent with the previous work. It is independent of the number of pages per block.

Now, Assumption \ref{binomial_assumption} is justified. Consider a certain block, at the very beginning of its block cycle with $x$ free pages and $N_{\mathsf{p}}-x$ valid pages. After $k_{\mathsf{e}}$ user writes, this block is again selected for garbage collection. Thus, all the valid pages in the beginning of the block cycle have the same probability $p_{\mathsf{ie}}$ to be invalid at the end of the block cycle. The free pages in the beginning of the block cycle are written consecutively and each of them experience one less user write than the previous page. Practical flash memory has a very large number of blocks so that $Tx\gg 1$. Thus, it is reasonable to assume that each page in a block has the same probability $p_{\mathsf{ie}}$ to be invalid when selected for garbage collection. 

\section{Simulation}\label{simulation}
A flash memory simulator was written and simulations were run to justify the assumptions and verify the validity of equation (\ref{final}). All simulations below start with an empty flash memory.

\begin{figure}[t]
\centering
\scalebox{0.5}{\includegraphics{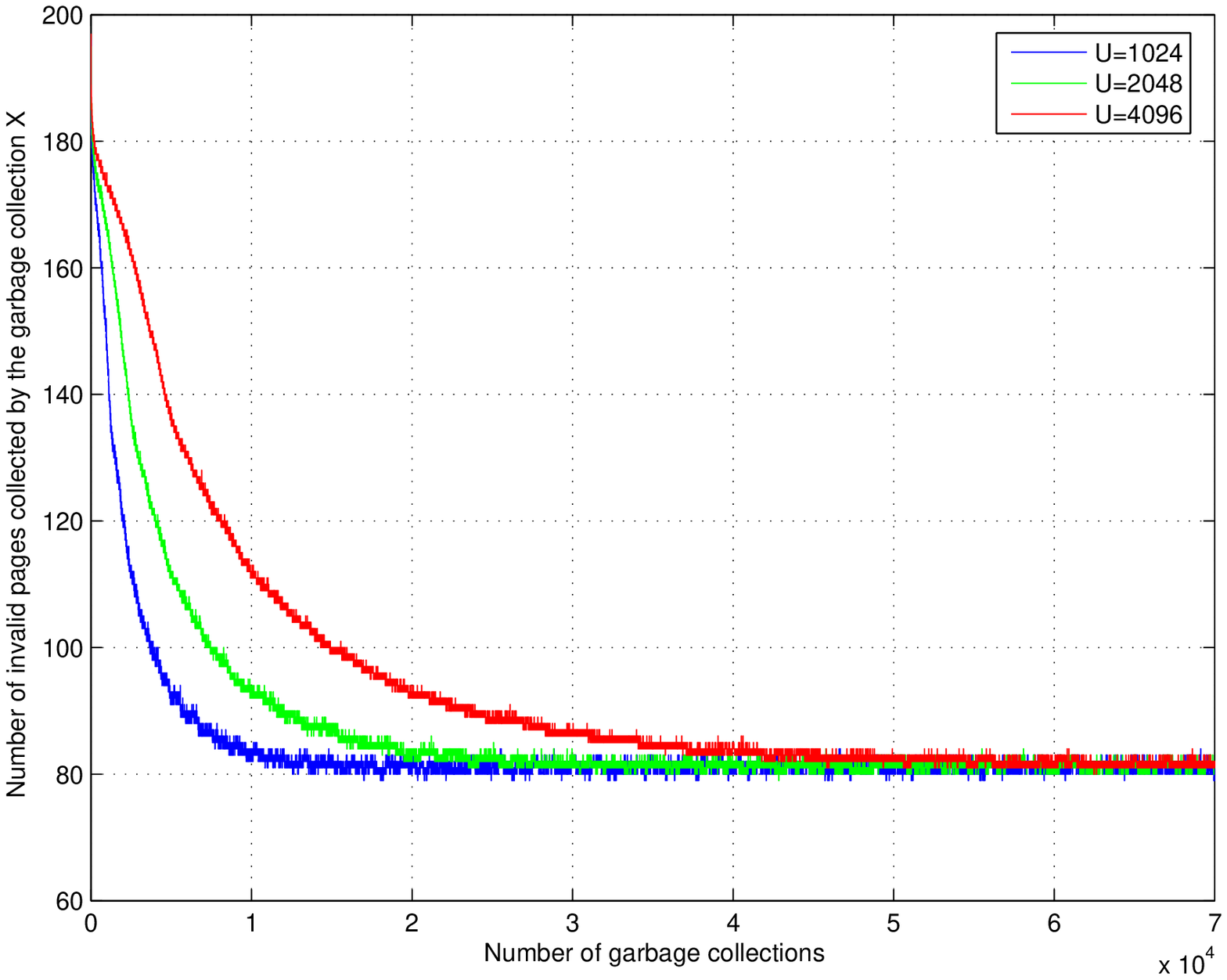}}
\caption{Evolution of the number of invalid pages freed by garbage collection}
\label{justification_assumption11}
\end{figure}
\begin{figure}[b]
\centering
\scalebox{0.5}{\includegraphics{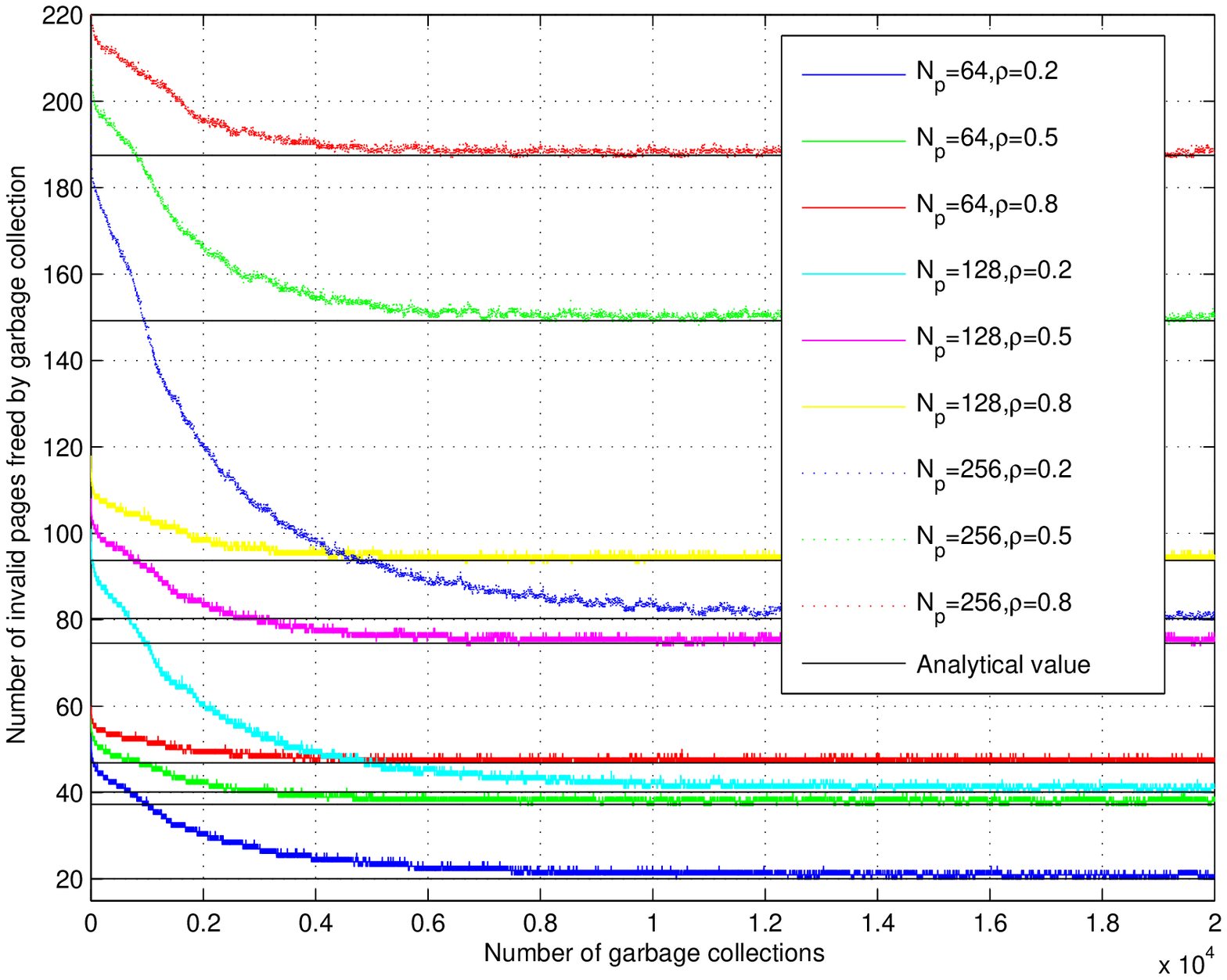}}
\caption{Evolution of the number of invalid pages freed by garbage collection}
\label{justification_assumption12}
\end{figure}

First, Assumption \ref{stationary_assumption} is justified. Simulations were run with different combinations of $U$, $N_\mathsf{p}$ and $\rho$. Fig. \ref{justification_assumption11} shows that when $\rho=0.20$ and $N_p=256$, $X$, the number of invalid pages freed by garbage collection, evolves into a stationary state after a sufficiently large number of garbage collections. It also shows that $X$ converges to the same value, which is close to the analytical result 80.31 given by equation (\ref{practical_x}), regardless of the value of $U$. This complies with equation (\ref{practical_x}) that $\widehat{x}$ is independent of $U$. Fig. \ref{justification_assumption12} shows that $X$ evolves into a stationary state regardless of the value of $N_\mathsf{p}$ and $\rho$ as well when the number of user writes is sufficiently large. Therefore, after a sufficiently large number of user writes, the average value of $X$ always converges to a constant $x$. It also closely approaches the analytical value.

\begin{figure}[t]
\centering
\scalebox{0.5}{\includegraphics{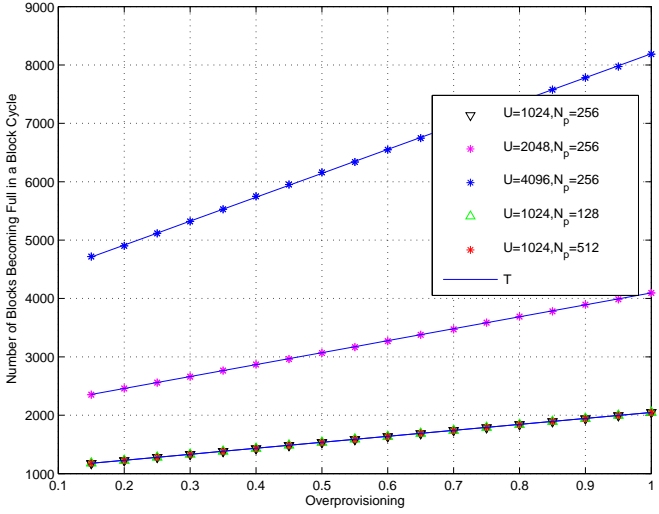}}
\caption{Approximation of number of blocks becoming full in a block cycle using T}
\label{TX}
\end{figure}

Fig. \ref{TX} shows that for different values of $U$, $N_p$ and $\rho$, $T$ is always a good approximation of the number of blocks becoming full in a block cycle. This justifies Assumption \ref{T_assumption}.

\begin{figure}[b]
\centering
\scalebox{0.5}{\includegraphics{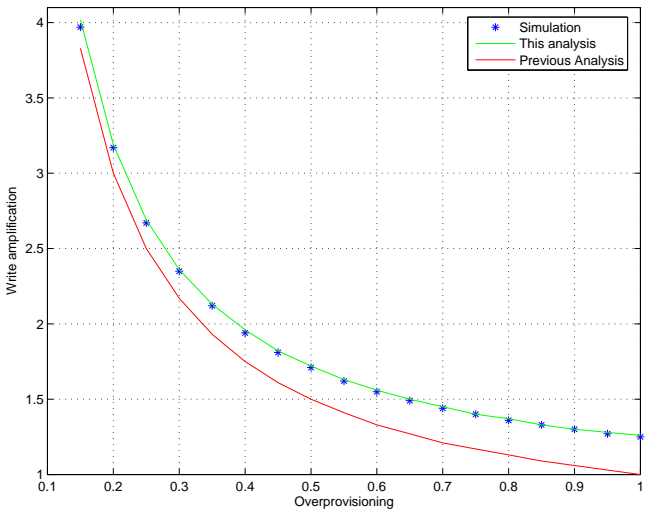}}
\caption{Comparison of results}
\label{fig:comparison}
\end{figure}

Finally, equation (\ref{final}) is evaluated and compared to the previous work equation (\ref{previous_work}). From equation (\ref{final}), write amplification is independent of $U$ and $N_{\mathsf{p}}$. Thus, comparison is made only for different $\rho$ values. Simulation is run using $N_p=256$, $U=1024$, and $\rho$  varies from 0.15 to 1.00. The comparison of results from simulation, this analysis (equation (\ref{final})) and previous work (equation (\ref{previous_work})) is shown in Fig. \ref{fig:comparison} and Table \ref{table:comparison}. This shows that, the result of equation (\ref{final}) is much closer to the simulation value than the result of previous work. For example, when $\rho=0.30$ which is a commonly adopted value in practical flash memories, equation (\ref{final}) predicts $\widehat{A}=2.36$ whereas equation (\ref{previous_work}) gives $A_{\mathsf{pre}}=2.17$, when the actual value is $2.35$.

\begin{table}
\begin{center}
\caption{Comparison of Results}
\begin{tabular}{| c | c | c | c |}
\hline
$\rho$ & Simulation & This analysis & Previous analysis\\
\hline
     0.15      &		3.97	  	&		4.02	&3.83\\
\hline
      0.20     &		3.17		&		3.19	&3.00\\
\hline
      0.25     &			2.67	&		2.69	&2.50	 \\
\hline
      0.30     &		2.35		&		2.36	&2.17\\
\hline
      0.35     &			2.12	&	 2.13		&1.93\\
\hline
       0.40    &		  1.94   	&	 1.96		&1.75\\
\hline
     0.45      &		1.81 	& 	 1.82	     &1.61	\\
\hline
      0.50     &			1.71	&		1.72	& 1.50\\
\hline
     0.55      &			1.62	&		1.63	&1.41\\
\hline
     0.60      &		 1.55		&		1.56	&1.33\\
\hline
     0.65      &			1.49	&		1.50	&1.27\\
\hline
      0.70     &		1.44		&		1.45	& 1.21\\
\hline
     0.75      &		1.40	& 			 1.40	&  1.17\\
\hline
      0.80     &		1.36		&			1.37		&1.13	\\
\hline
     0.85      &		 1.33		&			1.33			&1.09\\
\hline
      0.90     &			1.30	&		 1.30			&1.06 \\
\hline
      0.95     &			1.27	&			1.28		& 1.03\\
\hline
         1.00  &			1.25	&			1.26		&1.00\\
\hline
\end{tabular}
\label{table:comparison}
\end{center}
\end{table}

\end{document}